# Constraints on CP violating four-fermion interactions*

Xiao-Gang He and Bruce McKellar

*School of Physics, University of Melbourne*

*Parkville, Vic. 3052, Australia*

(April 1996)

## Abstract

It has been shown that CP violating electron-nucleon and nucleon-nucleon interactions can induce atomic electric dipole moments and are therefore constrained from experimental data. We show that using the experimental upper bounds on neutron and electron electric dipole moments, one can also obtain constraints, in some cases better ones, on these interactions. In addition stringent constraints can also be obtained for muon-quark and tauon-quark four-fermion CP violating interactions, which cannot be constrained from atomic electric dipole moment experiments.

*Work supported in part by Australian Research Council.

A non-zero electric dipole moment (EDM) of an atom signals CP violation. There have been extensive studies of atomic EDM in the past several years both experimentally [1–3] and theoretically [4–8]. A non-zero atomic EDM can be induced by several CP violating interactions: electron and nucleon EDM's, CP violating nucleon-nucleon and electron-nucleon interactions, and etc. So far only upper bounds on atomic EDM have been obtained; these bounds constrain the underlying CP violating interactions. Here we concentrate on constraints on CP violating electron-nucleon and nucleon-nucleon interactions and their generalization to lepton-quark and quark-quark four-fermion interactions. The CP violating electron-nucleon and nucleon-nucleon interactions can be written as:

$$C_S \frac{G_F}{\sqrt{2}} i\bar{N}N\bar{e}\gamma_5 e\ ,\quad C_P \frac{G_F}{\sqrt{2}} i\bar{N}\gamma_5 N\bar{e}e\ ,\quad C_T \frac{G_F}{\sqrt{2}} i\bar{N}\sigma_{\mu\nu}N\bar{e}\sigma^{\mu\nu}\gamma_5 e\ ,$$
$$\eta_{NN'} \frac{G_F}{\sqrt{2}} i\bar{N}\gamma_5 N\bar{N}'N'\ ,\quad \eta'_{NN'} \frac{G_F}{\sqrt{2}} i\bar{N}\sigma_{\mu\nu}N\bar{N}'\sigma^{\mu\nu}\gamma_5 N'\ , \tag{1}$$

where $N$ and $N'$ can be $n$ or $p$. The best experimental constraints on these interactions come from bound on the EDM of $^{199}Hg$ [2,6] with $|C_S| < 1\times 10^{-6}$ and $|C_T| < 2\times 10^{-8}$. The constraint on $C_P$ is about one order of magnitude weaker than that on $C_S$. The parameter $|\eta|$ is constrained to be less than 0.1 [4,5]. No constraint has been obtained for $\eta'$.

The CP violating interactions enumerated in eq.(1) are generated by some underlying CP violating interaction at the leptons, quarks and their interactions. In particular these interactions can be generated by four-fermion lepton-quark and quark-quark interactions. In this note we point out that CP violating interactions at this level can also be constrained by the experimental bounds on the electron EDM ($d_e < 10^{-26}$ ecm [3]) and neutron EDM ($d_n < 10^{-25}$ecm [9]). In this paper we determine these bounds on the underlying four-fermion lepton-quark and quark-quark interactions, and the implied constraints on the parameters of the interactions of equation (1). It will be seen that this analysis leads to improved bounds on $\eta_{pn}$.

At the elementary particle level, the interactions in eq.(1) are related to the following lepton-quark and quark-quark interactions

$$\bar{e}\gamma_5 e\bar{q}q\ ,\quad \bar{e}e\bar{q}\gamma_5 q\ ,\quad \bar{e}\sigma_{\mu\nu}e\bar{q}\sigma^{\mu\nu}\gamma_5 q\ ,$$

$$\bar{q}q\bar{q}'\gamma_5 q' \ , \ \ \bar{q}\sigma_{\mu\nu}q\bar{q}'\sigma^{\mu\nu}\gamma_5 q' \ , \tag{2}$$

by the relations: $\langle N|m_q\bar{q}q|N\rangle = \Delta_{Nq}\bar{N}N$, $\langle N|m_q\bar{q}\gamma_5 q|N\rangle = \Delta'_{Nq}\bar{N}\gamma_5 N$, and the non-relativistic valence quark model approximation, $\langle n|\bar{u}\sigma_{\mu\nu}u|n\rangle = (-1/3)\bar{n}\sigma_{\mu\nu}n$, $\langle n|\bar{d}\sigma_{\mu\nu}d|n\rangle = (4/3)\bar{n}\sigma_{\mu\nu}n$. Here $\Delta_{Nq}$ can be determined from experimental data and calculations of the nucleon mass shift due to SU(3)-breaking quark masses, and $\Delta'_{Nq}$ can be obtained from polarized proton and neutron experimental data [10]. Similarly, quark-quark interactions can be related to nucleon-nucleon interactions. We will confine ourselves to flavour conserving interactions, that is, we do not consider interactions which convert different lepton or quark generation. Flavour conserving CP violating interactions are very small (at higher than two loop level order) in the Standard Model (SM) [7,8]. If relatively large interactions are shown to exist experimentally, they are signals of physics beyond the SM.

One may wonder how the lepton-quark and quark-quark four-fermion interactions may be generated beyond the SM. The $C_{S,P}$ type of interactions can be generated, for example, by exchanging neutral Higgs scalars or leptoquark scalars at the tree level [7,8]. Interactions of the $C_T$ type can be generated by exchanging leptoquark scalars at tree level, and the $\eta$ interaction can be generated by exchanging a di-quark scalar at the tree level [7,8]. In the following we will not restrict ourselves to particular models, but use an effective Lagrangian approach to analyze possible constraints on the strength of these interactions [11,12]. We assume that any new physics introduced beyond the SM, say at a scale $\Lambda > m_Z$, which may have a gauge symmetry different to that at the scale $\Lambda$, is such that symmetry breaking between $\Lambda$ and $m_Z$ gives the SM gauge symmetry as a residual symmetry at the electroweak scale. We will first analyze all possible lepton-quark and quark-quark interactions which respect the SM gauge symmetry $SU(3)_C \times SU(2)_L \times U(2)_Y$ and identify the relevant CP violating interactions, and then constrain these interactions using experimental data.

The left-, and right-handed quarks $q^i_L$, $u^i_R$ , $d^i_R$, and left-, and right-handed leptons $l^i_L$, $e^i_R$ transform under the SM gauge group $SU(3)_C \times SU(2)_L \times U(2)_Y$ as:

$$q^i_L(3,\ 2,\ 1/3), \quad u^i_R(3,\ 1,\ 4/3), \ \ d^i_R(3,\ 1,\ -2/3)$$

$$l^i_L(1,\ 2,\ -1),\ \ e^i_R(1,\ 1,\ -2)\ , \tag{3}$$

where i is the generation index.

There are four different chiral structures for four-fermion interactions, $\bar{L}L\bar{L}L$, $\bar{R}R\bar{R}R$, $\bar{L}R\bar{R}L$, and $\bar{L}R\bar{L}R$ with appropriate Lorentz structures associated with them. Here $L$, and $R$ indicate left-, and right-handed fermions. We find that CP violation can only occur in operators with the $\bar{L}R\bar{R}L$ and $\bar{L}R\bar{L}R$ structure. Among these operators only the following ones contribute to fermion EDMs at one loop level,

$$\begin{aligned} &O_1^{ijkl} = \bar{l}^i_L \sigma_{\mu\nu} e^j_R \bar{q}^k_L \sigma^{\mu\nu} u^l_R\ , \quad O_2^{ijkl} = \bar{q}^i_L u^j_R \bar{q}^k_L d^l_R\ , \quad O_3^{ijkl} = \bar{q}^i_L \sigma_{\mu\nu} u^j_R \bar{q}^k_L \sigma^{\mu\nu} d^l_R\ ,\\ &O_4^{ijkl} = \bar{q}^i_L T^a u^j_R \bar{q}^k_L T^a d^l_R\ , \quad O_5^{ijkl} = \bar{q}^i_L \sigma_{\mu\nu} T^a u^j_R \bar{q}^k_L \sigma^{\mu\nu} T^a d^l_R\ , \end{aligned} \tag{4}$$

where $T^a$ is the $SU(3)_C$ generator and is normalized as $\mathrm{Tr}(T^aT^b) = \delta^{ab}/2$. The operators $O_{1,3,5}$ are not listed in Ref. [11]. They are independent of $O_2$ and $O_4$ and should be considered in calculating the EDM. All the operators listed above have dimension six, and are expected to be generated at the high energy scale $\Lambda$ by some unknown physics. Flavour conservation requires: $i = j$, $k = l$ or $i = l$, $j = k$. For $O_1$, only $i = j$, $k = l$ is allowed. To reflect the scale at which these operators are generated and their dimensionality, we parameterize the effective Lagrangian as

$$\begin{aligned} L = &\frac{\lambda_1^{ij}}{\Lambda^2} \bar{l}^i_L \sigma_{\mu\nu} e^i_R \bar{q}^j_L \sigma^{\mu\nu} u^j_R + \frac{\lambda_2^{ij}}{\Lambda^2} \bar{q}^i_L u^j_R \bar{q}^j_L d^i_R + \frac{\lambda_2'^{ij}}{\Lambda^2} \bar{q}^i_L u^i_R \bar{q}^j_L d^j_R + \frac{\lambda_3^{ij}}{\Lambda^2} \bar{q}^i_L \sigma_{\mu\nu} u^j_R \bar{q}^j_L \sigma^{\mu\nu} d^i_R\\ &+ \frac{\lambda_3'^{ij}}{\Lambda^2} \bar{q}^i_L \sigma_{\mu\nu} u^i_R \bar{q}^j_L \sigma^{\mu\nu} d^j_R + \frac{\lambda_4^{ij}}{\Lambda^2} \bar{q}^i_L T^a u^j_R \bar{q}^j_L T^a d^i_R + \frac{\lambda_4'^{ij}}{\Lambda^2} \bar{q}^i_L T^a u^i_R \bar{q}^j_L T^a d^j_R\\ &+ \frac{\lambda_5^{ij}}{\Lambda^2} \bar{q}^i_L \sigma_{\mu\nu} T^a u^j_R \bar{q}^j_L \sigma^{\mu\nu} T^a d^i_R + \frac{\lambda_5'^{ij}}{\Lambda^2} \bar{q}^i_L \sigma_{\mu\nu} T^a u^i_R \bar{q}^j_L \sigma^{\mu\nu} T^a d^j_R + h.c.\ , \end{aligned} \tag{5}$$

where for $\lambda'^{ij}$, we impose the restriction $i \neq j$ to avoid double counting, since the $\lambda^{ii}$ and $\lambda'^{ii}$ interactins are identical. To compare with the interaction strength of the standard Fermi interaction, $\lambda^{ij}/\Lambda^2$ is some times conveniently written as $C^{ij}G_F/\sqrt{2}$ in the literature. The dimensionless parameter $C^{ij}$ indicates the relative strength of the new interaction. We prefer to use the parameterization in eq. (5) in our calculations to keep track of the energy scale. We will first use experimental data to obtain bounds on the parameters $\lambda^{ij}$, and then convert these bounds into the parameters $C_{S,T}$ and $\eta$.

The Feynman diagram responsible for the fermion EDM at one loop level is shown in figure 1. ¿From this we obtain, for the EDM of the leptons and the $u^i$ quark,

$$
\begin{aligned}
d_{l^i}(\lambda_1^{ij}) &= \frac{N}{12\pi^2} e\mathrm{Im}(\lambda_1^{ij}) m_{u^j} \frac{1}{\Lambda^2} \ln\frac{\Lambda^2}{m_{u^j}^2} , \\
d_{u^j}(\lambda_1^{ij}) &= -\frac{1}{8\pi^2} e\mathrm{Im}(\lambda_1^{ij}) m_{l^i} \frac{1}{\Lambda^2} \ln\frac{\Lambda^2}{m_{l^i}^2} , \\
d_{u^i}(\lambda_2^{ji}) &= \frac{1}{8}\frac{1}{8\pi^2} eQ_d\mathrm{Im}(\lambda_2^{ji}) m_{d^j} \frac{1}{\Lambda^2} \ln\frac{\Lambda^2}{m_{d^j}^2} , \\
d_{u^i}(\lambda_3^{ji}) &= \frac{2N\delta_{i1}\delta_{j1}-1}{2}\frac{1}{8\pi^2} eQ_d\mathrm{Im}(\lambda_3^{ji}) m_{d^j} \frac{1}{\Lambda^2} \ln\frac{\Lambda^2}{m_{d^j}^2} , \\
d_{u^i}(\lambda_3'^{ij}) &= \frac{2N-1}{2}\frac{1}{8\pi^2} eQ_d\mathrm{Im}(\lambda_3'^{ij}) m_{d^j} \frac{1}{\Lambda^2} \ln\frac{\Lambda^2}{m_{d^j}^2} , \\
d_{u^i}(\lambda_4^{ji}) &= \frac{N^2-1}{16N}\frac{1}{8\pi^2} eQ_d\mathrm{Im}(\lambda_4^{ji}) m_{d^j} \frac{1}{\Lambda^2} \ln\frac{\Lambda^2}{m_{d^j}^2} , \\
d_{u^i}(\lambda_5^{ji}) &= -\frac{N^2-1}{4N}\frac{1}{8\pi^2} eQ_d\mathrm{Im}(\lambda_5^{ji}) m_{d^j} \frac{1}{\Lambda^2} \ln\frac{\Lambda^2}{m_{d^j}^2} ,
\end{aligned} \tag{6}
$$

where $N = 3$ is the number of colours. Here we have taken the cut-off in the loop integral to be the same as the scale $\Lambda$. The operators $O_{2,3,4,5}$ also induce an EDM of the $d^j$ quark. The $d_{d^j}$ are obtained by replacing $d^j$, $u^i$ and $Q_d$ by $u^i$, $d^j$ and $Q_u$ in the above equations. We use the valence quark model to relate the neutron EDM to the quark EDM with: $d_n = (4/3)d_d - (1/3)d_u$. We note that not all operators of eq.(4) contribute to the fermion EDM. Therefore the EDM bounds will constrain only some of the parameters of the effective Lagrangian.

The results are logrithmically divergent. This is a good indication that the results are reliable. In fact if leptoquark or di-quark exchange is responsible for the EDM, the scale $\Lambda$ is just their masses [7,8]. The operators discussed here may indeed be generated by the exchange of some heavy particles at energy scale $\Lambda$. Several operators with $\bar{L}R\bar{R}L$ and $\bar{L}R\bar{L}R$ chiral structures can induce phases in the determinants of the quark mass matrices and therefore a non-zero strong QCD CP violating $\theta$ term. If one naively uses the constraint on $\theta$, one would obtain very stringent bounds on the strength of these operators. However, the phases calculated in this way are quadratically divergent. We regard this as unsatisfactory, and feel that until one has a better understanding of the physics at the scale $\Lambda$, these results should not be used to give bounds on the parameters of eq.(5).

In principle one can use the EDM bounds on the neutron, electron, muon and tauon to constrain the parameters. We find that only the neutron and electron EDM's can provide significant constraints. Assuming that there is no accidental cancellation among different contributions, we obtain constraints given in Table I.

Using experimental bounds on $C_{S,T}$ and $\eta$, we can also obtain constraints on the parameters $\lambda^{ij}$. These constraints, in most cases, are much less stringent than those obtained in Table I. To compare with constraints on the parameters from direct atomic EDM measurements, we write the constraints obtained in Table I in terms of the parameters $C^n_T$, $\eta_{pn}$. We use the factorization approximation and the non-relativistic valence quark model to estimate the matrix elements. We have,

$$\begin{aligned}
&\langle n|\,\bar e\sigma_{\mu\nu}e\bar u\sigma^{\mu\nu}\gamma_5 u|n\rangle = -\frac{1}{3}\bar e\sigma_{\mu\nu}e\bar n\sigma^{\mu\nu}\gamma_5 n\ ,\\
&\langle pn|O_2^{1111}|pn\rangle|_{CP} = \frac{1}{4}\frac{2N+1}{2N}[\langle n|\bar u^1 u^1|n\rangle\langle p|\bar d^1\gamma_5 d^1|p\rangle\\
&+\ \langle n|\bar d^1 d^1|n\rangle\langle p|\bar u^1\gamma_5 u^1|p\rangle] = \frac{1}{4}\frac{2N+1}{2N}(\Delta_{nu^1}\Delta'_{pd^1}+\Delta_{nd^1}\Delta'_{pu^1})\bar n n\bar p\gamma_5 p\ ,\\
&\langle pn|O_2^{ijji}|pn\rangle|_{CP} = \frac{1}{8N}[\langle n|\bar u^i u^i|n\rangle\langle p|\bar d^j\gamma_5 d^j|p\rangle\\
&+\ \langle n|\bar d^j d^j|n\rangle\langle p|\bar u^i\gamma_5 u^i|p\rangle] = \frac{1}{8N}(\Delta_{nu^i}\Delta'_{pd^j}+\Delta_{nd^i}\Delta'_{pu^i})\bar n n\bar p\gamma_5 p\ ,\\
&\langle np|O_3^{1111}|np\rangle|_{CP} = \frac{12}{2N+1}\langle np|O_2^{1111}|np\rangle|_{CP}\ ,\\
&\langle np|O_3^{ijji}|np\rangle|_{CP} = 12\langle np|O_2^{ijji}|np\rangle|_{CP}\ ,\\
&\langle np|O_4^{1111}|np\rangle|_{CP} = \frac{N^2-1}{2N(2N+1)}\langle np|O_2^{1111}|np\rangle|_{CP}\ ,\\
&\langle np|O_4^{ijji}|np\rangle|_{CP} = \frac{N^2-1}{2N}\langle np|O_2^{ijji}|np\rangle|_{CP}\ ,\\
&\langle np|O_5^{1111}|np\rangle|_{CP} = \frac{6(N^2-1)}{N(2N+1)}\langle np|O_2^{1111}|np\rangle|_{CP}\ ,\\
&\langle np|O_5^{ijji}|np\rangle|_{CP} = \frac{6(N^2-1)}{N}\langle np|O_2^{ijji}|np\rangle|_{CP}\ .
\end{aligned}\tag{7}$$

In our numerical evaluations, we will use the values $m_u = 4.2$ MeV, $m_d = 7.5$ MeV, and $m_s = 150$ MeV:

$$\Delta_{nu} = \frac{18\mathrm{MeV}}{m_u}\ ,\quad \Delta_{nd} = \frac{18\mathrm{MeV}}{m_d}\ ,\ \Delta_{ns} = \frac{247\mathrm{MeV}}{m_s}\ ,$$

$$\Delta_{nh} = \frac{48\text{MeV}}{m_h} \ , \ \ \Delta'_{pu} = \frac{432\text{MeV}}{m_u} \ , \ \ \Delta'_{d} = -\frac{419\text{MeV}}{m_d} \ ,$$
$$\Delta'_{ns} = -\frac{165\text{MeV}}{m_s} \ , \ \ \Delta'_{ph} = -\frac{63\text{MeV}}{m_h} \ ,$$

where $h$ indicates a heavy quark [10]. The operators with coefficients $\lambda'_3$ do not contribute to $\eta_{pn}$. The constraints on $C^n_T$ and $\eta_{pn}$ are shown in Table II.

The constraints on $C^n_T$ are much weaker than those obtained from the upper bound on the atomic EDM. However, the constraints on $\eta_{pn}$ in most cases are much better than those obtained from atomic EDM. In addition, we obtained constraints on muon-quark and tauon-quarks interactions which cannot be obtained from atomic electric dipole moment measurements.

# TABLES

TABLE I. The upper bounds on $|\mathrm{Im}\lambda^{ij}|$ from the electric dipole moments of electron and neutron for $\Lambda = 1\mathrm{TeV}$.

| | | | | | |
|---|---|---|---|---|---|
| $d_e$ | | | | | |
| | $\lvert\mathrm{Im}(\lambda_1^{eu})\rvert$ | $\lvert\mathrm{Im}(\lambda_1^{ec})\rvert$ | $\lvert\mathrm{Im}(\lambda_1^{et})\rvert$ | | |
| | $2.0 \times 10^{-4}$ | $1.2 \times 10^{-6}$ | $3.3 \times 10^{-8}$ | | |
| $d_n$ | | | | | |
| | $\lvert\mathrm{Im}(\lambda_1^{eu})\rvert$ | $\lvert\mathrm{Im}(\lambda_1^{\mu u})\rvert$ | $\lvert\mathrm{Im}(\lambda_1^{\mu u})\rvert$ | | |
| | $8.3 \times 10^{-2}$ | $6.5 \times 10^{-4}$ | $5.3 \times 10^{-5}$ | | |
| | $\lvert\mathrm{Im}(\lambda_2^{du})\rvert$ | $\lvert\mathrm{Im}(\lambda_2^{dc})\rvert$ | $\lvert\mathrm{Im}(\lambda_2^{dt})\rvert$ | $\lvert\mathrm{Im}(\lambda_2^{su})\rvert$ | $\lvert\mathrm{Im}(\lambda_2^{bu})\rvert$ |
| | $3.7 \times 10^{-2}$ | $2.1 \times 10^{-4}$ | $5.9 \times 10^{-6}$ | $1.0 \times 10^{-2}$ | $5.4 \times 10^{-4}$ |
| | $\lvert\mathrm{Im}(\lambda_3^{du})\rvert$ | $\lvert\mathrm{Im}(\lambda_3^{dc})\rvert$ | $\lvert\mathrm{Im}(\lambda_3^{dt})\rvert$ | $\lvert\mathrm{Im}(\lambda_3^{su})\rvert$ | $\lvert\mathrm{Im}(\lambda_3^{bu})\rvert$ |
| | $1.6 \times 10^{-2}$ | $5.3 \times 10^{-5}$ | $1.5 \times 10^{-6}$ | $2.5 \times 10^{-3}$ | $1.4 \times 10^{-4}$ |
| | | $\lvert\mathrm{Im}(\lambda_3'^{cd})\rvert$ | $\lvert\mathrm{Im}(\lambda_3'^{td})\rvert$ | $\lvert\mathrm{Im}(\lambda_3'^{us})\rvert$ | $\lvert\mathrm{Im}(\lambda_3'^{ub})\rvert$ |
| | | $8.8 \times 10^{-6}$ | $2.5 \times 10^{-7}$ | $4.1 \times 10^{-4}$ | $2.2 \times 10^{-5}$ |
| | $\lvert\mathrm{Im}(\lambda_4^{du})\rvert$ | $\lvert\mathrm{Im}(\lambda_4^{dc})\rvert$ | $\lvert\mathrm{Im}(\lambda_4^{dt})\rvert$ | $\lvert\mathrm{Im}(\lambda_4^{su})\rvert$ | $\lvert\mathrm{Im}(\lambda_4^{bu})\rvert$ |
| | $2.8 \times 10^{-2}$ | $1.6 \times 10^{-4}$ | $4.4 \times 10^{-6}$ | $0.75 \times 10^{-2}$ | $4.1 \times 10^{-4}$ |
| | $\lvert\mathrm{Im}(\lambda_5^{du})\rvert$ | $\lvert\mathrm{Im}(\lambda_5^{dc})\rvert$ | $\lvert\mathrm{Im}(\lambda_5^{dt})\rvert$ | $\lvert\mathrm{Im}(\lambda_5^{su})\rvert$ | $\lvert\mathrm{Im}(\lambda_5^{bu})\rvert$ |
| | $0.69 \times 10^{-2}$ | $3.9 \times 10^{-4}$ | $1.1 \times 10^{-6}$ | $1.9 \times 10^{-3}$ | $1.0 \times 10^{-4}$ |

TABLE II. The upper bounds on $C_T^n$ and $\eta_{pn}$ with $\Lambda_5 = 1$ TeV.

| | | | | | |
|---|---|---|---|---|---|
| $d_e$ | | | | | |
| | $\lvert C_T^n(\lambda_1^{eu})\rvert$ | | | | |
| | $0.8 \times 10^{-8}$ | | | | |
| $d_n$ | | | | | |
| | $C_T^n(\lambda_1^{eu})\rvert$ | | | | |
| | $3.4 \times 10^{-3}$ | | | | |
| | $\lvert\eta_{pn}(\lambda_2^{du})\rvert$ | $\lvert\eta_{pn}(\lambda_2^{dc})\rvert$ | $\lvert\eta_{pn}(\lambda_2^{dt})\rvert$ | $\lvert\eta_{pn}(\lambda_2^{su})\rvert$ | $\lvert\eta_{pn}(\lambda_3^{bu})\rvert$ |
| | 0.28 | $8.3 \times 10^{-6}$ | $1.8 \times 10^{-9}$ | $1.7 \times 10^{-2}$ | $5.2 \times 10^{-6}$ |
| | $\lvert\eta_{pn}(\lambda_3^{du})\rvert$ | $\lvert\eta_{pn}(\lambda_3^{dc})\rvert$ | $\lvert\eta_{pn}(\lambda_3^{dt})\rvert$ | $\lvert\eta_{pn}(\lambda_3^{su})\rvert$ | $\lvert\eta_{pn}(\lambda_3^{bu})\rvert$ |
| | 0.21 | $2.5 \times 10^{-5}$ | $5.5 \times 10^{-9}$ | $5.1 \times 10^{-2}$ | $1.6 \times 10^{-5}$ |
| | $\lvert\eta_{pn}(\lambda_4^{du})\rvert$ | $\lvert\eta_{pn}(\lambda_4^{dc})\rvert$ | $\lvert\eta_{pn}(\lambda_4^{dt})\rvert$ | $\lvert\eta_{pn}(\lambda_4^{su})\rvert$ | $\lvert\eta_{pn}(\lambda_4^{bu})\rvert$ |
| | $4.0 \times 10^{-2}$ | $8.5 \times 10^{-6}$ | $1.8 \times 10^{-9}$ | $1.7 \times 10^{-2}$ | $5.3 \times 10^{-6}$ |
| | $\lvert\eta_{pn}(\lambda_5^{du})\rvert$ | $\lvert\eta_{pn}(\lambda_5^{dc})\rvert$ | $\lvert\eta_{pn}(\lambda_5^{dt})\rvert$ | $\lvert\eta_{pn}(\lambda_5^{su})\rvert$ | $\lvert\eta_{pn}(\lambda_5^{bu})\rvert$ |
| | 0.12 | $2.5 \times 10^{-5}$ | $5.4 \times 10^{-9}$ | $5.1 \times 10^{-2}$ | $1.5 \times 10^{-5}$ |

## FIGURES

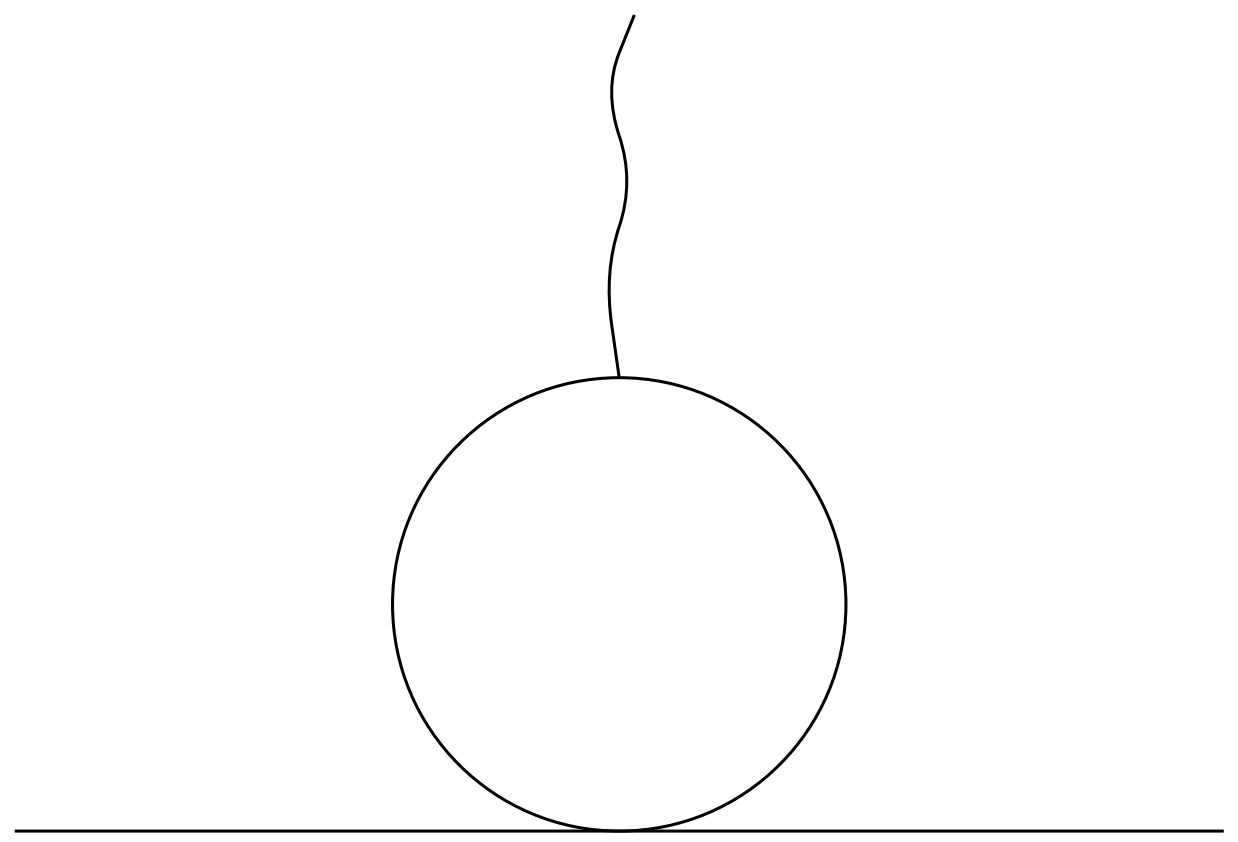

FIG. 1. One loop Feynman diagram for fermion EDM.